# Determining the full transformation relations in the transformation method

**Jin Hu[1][*], and Xiang-Yang Lu[1,2]**

[1] School of Information and Electronics, Beijing Institute of Technology, Beijing, China
[2] School of Electronic and Information Engineering, Zhongyuan University of Technology, Zhengzhou, China
*corresponding author, E-mail: `bithj@bit.edu.cn`

**Abstract**

Transformation method provides an efficient way to control wave propagation by materials. However, the degree to which this transformation concept can be applied to other physical phenomena remains an open question. Recently, Hu et al. presented a general framework for determining the transformation relations of the physical quantities in arbitrary partial differential equation (PDE) in its application scope according to the idea of transformation method. In this paper, we will review the main concepts of this general theory, particularly the difference between this idea and usual methods. The flexibility of this method is shown using an example.

## 1. Introduction

The transformation method proposed for electromagnetic (EM) waves [1-3] has many applications for controlling and manipulating electromagnetic fields with the help of electromagnetic metamaterials, such as cloaks [2,4]. The method can also be extended to acoustic waves for liquid materials [5,6], heat conduction [7] and matter waves [8]. The basic principle of the transformation method is based on the form-invariance of the corresponding equations under general spatial mapping. Then materials needed to distribute the fields in a prescribed manner can be constructed directly. Consider a physical process described on an initial space $\Omega$, the field **u** and material **C** are related together at a point **x** and time $t$ by a differential equation $F$ as

$$F(\mathbf{x}, t, \mathbf{u}(\mathbf{x},t), \mathbf{C}(\mathbf{x})) = 0, \quad \mathbf{x} \in \Omega. \tag{1}$$

The operator $F$ announces the certain physical mechanism between **C** and **u** in every point within $\Omega$. If $F$ is form-invariant under the mapping $\mathbf{x}' = \mathbf{x}'(\mathbf{x})$, which transforms the space $\Omega$ to $\Omega'$, i.e. there is

$$F(\mathbf{x}', t, \mathbf{u}'(\mathbf{x}',t), \mathbf{C}'(\mathbf{x}')) = 0, \quad \mathbf{x}' \in \Omega', \tag{2}$$

then the attached field **u** and materials **C** in $\Omega$ can also be mapped to $\Omega'$ as **u'** and **C'**. The idea of the transformation method is to carefully perform a specific mapping so that the transformed field **u'** in the transformed space follows a designed way, and **C'** then tells how to realize this function by material distribution. Hence, the transformation relations for field and material during a transformation, i.e.,

$$T_u : \mathbf{u} \mapsto \mathbf{u}', \quad T_C : \mathbf{C} \mapsto \mathbf{C}', \tag{3}$$

are essential to fulfill this method.

To obtain Eq. (3) for different equations, different groups have come up with different methods. Pendry et al. [2] rebuilt the Maxwell's equations in a small parallelepiped and found the transformation relations with the help of both the contravariant and covariant components of vectors. A similar method was used by Cummer et al. [9] to obtain the velocity transformation for scalar acoustic equations. Greenleaf et al. [10] analyzed the Helmholtz equation with respect to Riemannian metrics, and then the transformation relations could be obtained by comparison of the equations in Riemannian space and Euclidian space. Milton et al. [11] gave another concise mathematical formulation. They utilized tensor calculation to prove that for a special transformation of field and material the Maxwell's equations could retain their form. Norris [12] showed that the material transformations of generalized-acoustic equations were not uniquely defined based on the pentamode elastic material theory [13]. Interestingly, Leonhardt et al. [14] found that general relativity provides the recipe for calculating the required electromagnetic material properties. Recently, the so-called change-of-variable method which was developed by Milton et al. [11] has been extended by Vasquez et al. [15] and Norris et al. [16] to obtain the transformations of the elastodynamics equations. The main idea of the change-of-variable method is that one should first assume certain transformation relations for some physical quantities between **x** and **x'**, and then reestablish the whole equations with respect to **x'**, according to calculus rules. If the reestablished equations have the same form as the original one, then the equations are declared to be form-invariant, and the transformation relations Eq. (3) can be easily obtained by comparing the two sets of equations; otherwise, the equations are considered as form-variant and thus the transformation method can not be applied. By this method, the two groups achieved coincidental results for transformation elastodynamics, where the asymmetric elasticity tensor, which was first shown by Brun et al. [17], can be included as a special case.

The abovementioned methods are, generally speaking, based on the mathematical interpretation of form-invariance. In this interpretation, "form-invariance" is a pure mathematical property of an equation that one needs to verify. Recently, Hu et al. [18,19] and Chang et al. [20] have proposed another method to obtain Eq. (3). They were aware that the physical result of the space transformation is just converting the homogeneous (isotropous) materials to inhomogeneous (anisotropic) materials, so "form-invariance" is interpreted from the physical point of view. This interpretation argues that form-invariance is not a mathematical property of the governing equation, but the physics knowledge announcing the equation has the same form for both homogeneous and inhomogeneous materials. This knowledge is a foregone one, because any reasonable governing equation must have a clear answer for their application scope. The physical interpretation of form-invariance brings about conveniences in obtaining transformation relations during a transformation, however as it is totally different from the usual methods, the difference and relationship between these two ideas should be clarified. That is the objective of this paper.

The paper is arranged as follows: In Section 2, the change-of-variable method in the transformation method is discussed, and the ambiguity implied by this idea is pointed out, then in Section 3 the method based on the physical interpretation of form-invariance is discussed. A sample is given to show the flexibility of this method in Section 4, with a concluding summary in Section 5.

## 2. The change-of-variable method in transformation method

The change-of-variable method is a representative method to obtain Eq. (3) based on the mathematical interpretation of form-invariance. The main idea of this method has been briefly introduced in Section 1. In this section, we will give a more detailed introduction to this method in context of elastodynamics according to Ref. [11, 15-16], and the ambiguity of this method will be demonstrate subsequently.

The governing equations of elastodynamics are

$$\nabla \cdot \boldsymbol{\sigma} = -\omega^2 \rho \mathbf{u}, \quad \boldsymbol{\sigma} = \mathbf{C}\nabla \mathbf{u}, \quad (4)$$

where $\mathbf{u}$ denotes displacement vector, $\boldsymbol{\sigma}$ is 2-order stress tensor, $\mathbf{C}$ is 4-order elasticity tensor, $\rho$ is density and $\omega$ is the frequency. If one physical quantity transformation can be given first, then all the others can be obtained. For example, if one assumes that the displacement vector has this transformation [11]

$$\mathbf{u}' = (\mathbf{A}^T)^{-1}\mathbf{u}, \quad (5)$$

where $\mathbf{A}$ is the Jacobian transformation tensor with the components $A_{ij} = \partial x'_i/\partial x_j$, then the transformations of other physical quantities, $\boldsymbol{\sigma}$, $\mathbf{C}$, $\rho$ can be derived by submitting Eq. (5) to (4) and rewriting the equations with respect to $\mathbf{x}'$. The "test function"[11,15] or "Lagrangian density"[16] method can help with this derivation. One can find that the chain rule in calculus will lead to the form-variance of Eq. (4) in the new space $\Omega'$ [11]

$$\nabla' \cdot \boldsymbol{\sigma}' = \mathbf{D}'\nabla \mathbf{u}' - \omega^2 \boldsymbol{\rho}' \mathbf{u}', \quad \boldsymbol{\sigma}' = \mathbf{C}'\nabla'\mathbf{u}' + \mathbf{S}'\mathbf{u}', (6)$$

which are of the form of Willis' equations.

Realizing that the assumption of Eq. (5) is just a special case, Vasquez et al. [15] and Norris et al. [16] then propose a more general transformation of the displacement vector

$$\mathbf{u}' = \mathbf{B}^{-T}\mathbf{u}, \quad (7)$$

where $\mathbf{B}$ is an arbitrary invertible matrix-like transformation operator. As the arbitrary operator $\mathbf{B}$ is introduced, the transformations of quantities in Eq. (6) are nonunique [15,16]

$$C'_{ijkl} = J^{-1} A_{im} B_{jn} A_{kp} B_{lq} C_{mnpq},$$

$$S'_{ijk} = J^{-1} A_{im} B_{jn} \frac{\partial B_{kq}}{\partial x'_p} C_{mnpq}, \quad (8)$$

$$D'_{kij} = S'_{ijk},$$

$$\rho'_{ij} = J^{-1} B_{ik} B_{jk} \rho - J^{-1} \omega^{-2} \frac{\partial B_{in}}{\partial x'_m} \frac{\partial B_{jq}}{\partial x'_p} C_{mnpq},$$

where $J = \det \mathbf{A}$. Hence, it is obvious that

(a) If $\mathbf{C}'$ needs to be full symmetric, i.e., $C'_{ijkl} = C'_{jikl} = C'_{klij}$, then one has to set $\mathbf{B} = \mathbf{A}$, the transformed equations become Eq. (6), and this is the case by adopting the assumption of Eq. (5);

(b) If the transformed equations need to be kept in their original form, i.e., $\mathbf{D}' = \mathbf{S}' = 0$ in Eq. (6), then one has to set $\mathbf{B} = $ const, and thus $\mathbf{C}'$ loses it full symmetry because there may has $C'_{ijkl} \neq C'_{jikl}$. If $\mathbf{B} = \mathbf{I}$, then the result of Brun et al. [17] can be included as a special case.

So far it seems that the analyses are complete. However, the following question makes the above result debatable: although Eq. (7) is more general than Eq. (5), it is still a special case, so why the special result be treated as a general case and assert that Eq. (4) can not retain their form if the transformed elasticity tensor $\mathbf{C}'$ needs to be fully symmetric? The real general transformation of the displacement vector is

$$\mathbf{u}'(\mathbf{x}') = f(\mathbf{x}', \mathbf{x}, \mathbf{u}(\mathbf{x})), \quad (9)$$

where $f$ is an arbitrary continuous function, which can be much more complex than Eq. (7). In fact, as shown by Cevery [21], a time-harmonic solution of the elastodynamic equation (4) for an inhomogeneous medium can be written in the form of a vectorial ray series

$$\tilde{\mathbf{u}}(\mathbf{x}',t) = \left[\sum_{n=0}^{\infty} \frac{\mathbf{U}_n(\mathbf{x}')}{(-i\omega)^n}\right] \exp[-i\omega(t - T(\mathbf{x}'))], \quad (10)$$

and the system can be solved successively according to its recurrent character. Other literature about the displacement solutions of inhomogeneous elastodynamic can be found in Ref. [22]. This type of displacement vector can not be expressed by the simple transformation Eq. (7). For the isotropous homogeneous medium, we can assume the transformations of the material parameters from space $\Omega$ to $\Omega'$, which transforms the medium to inhomogeneous one

$$g(\mathbf{x}',\mathbf{x},\mathbf{C}):\mathbf{C} \mapsto \tilde{\mathbf{C}}(\mathbf{x}'),$$
$$h(\mathbf{x}',\mathbf{x},\rho):\rho \mapsto \tilde{\rho}(\mathbf{x}'), \qquad (11)$$

where $g$ and $h$ are given functions, and $\tilde{\mathbf{C}}$ is full symmetric by this transformation, then the displacement solution $\tilde{\mathbf{u}}$ can be obtained by Eq. (10) in space $\Omega'$. We have

$$\nabla \cdot \boldsymbol{\sigma} = -\omega^2 \rho \mathbf{u}, \quad \boldsymbol{\sigma} = \mathbf{C}\nabla \mathbf{u}, \quad \mathbf{x} \in \Omega$$
$$\nabla' \cdot \tilde{\boldsymbol{\sigma}} = -\omega^2 \tilde{\rho}\tilde{\mathbf{u}}, \quad \tilde{\boldsymbol{\sigma}} = \tilde{\mathbf{C}}\nabla'\tilde{\mathbf{u}}, \quad \mathbf{x}' \in \Omega' \qquad (12)$$

where $\mathbf{u}$ is the solution of the isotropous homogeneous medium in space $\Omega$, which can be of a harmonic plane wave form

$$\mathbf{u}(\mathbf{x}) = \mathbf{U}(\mathbf{x}) \exp(-i\omega(t - T(\mathbf{x}))). \qquad (13)$$

Thus, according to Eq. (12), there must exist point-to-point mapping between space $\Omega$ and $\Omega'$

$$\mathbf{C} \mapsto \tilde{\mathbf{C}}(\mathbf{x}'),$$
$$\rho \mapsto \tilde{\rho}(\mathbf{x}'), \qquad (14)$$
$$\mathbf{u}(\mathbf{x},t) \mapsto \tilde{\mathbf{u}}(\mathbf{x}',t).$$

Equations (14) clearly show that even if the elasticity tensor remains full symmetric, the elastodynamic equation (4) can be form-invariant for a general space mapping, providing the transformation of the displacement vector can be properly assumed. This result disagrees with assertions (a) and (b) in this section.

However, although the elastodynamic equation can, in principle, be form-invariant from a pure mathematical point of view, it does not indicate that Eq. (14) can be used in the framework of the transformation method to control the displacement field. This is because the inhomogeneous elastodynamic equations are equivalent to having non-physical sources term, thus the displacement solution Eq. (10) is not a smooth wave field, but a scattered wave field [21], which usually cannot satisfy engineering design requirements.

This fact demystifies the key point of the transformation method: the method does not concern *whether* the governing equations can be mapped (form-invariance) or not, instead, it concerns *what kind of* mapping (form-invariance) the equations have, or the characters of the transformation relations Eq. (3).

For high frequency elastic waves ($\omega \gg 1$), one can only use the zeroth-order term in the series of Eq. (10)

$$\tilde{\mathbf{u}}_0(\mathbf{x}',t) = \mathbf{U}_0(\mathbf{x}')\exp[-i\omega(t - T(\mathbf{x}')]. \qquad (15)$$

In this approximation, Eq. (10) is reduced to the form of plane wave form Eq. (13), which presents a smooth wave field, and can be used in the frame work of the transformation method. The local linear transformation must exist between the displacement vectors $\mathbf{u}$ and $\tilde{\mathbf{u}}_0$ because of their identical forms. This condition is also true for electromagnetic and acoustic waves: if the local linear transformation can be found, then Eq. (3) can be used in transformation method. However, this "local linear" restriction can not be naturally introduced into the change-of-variable method, because this restriction is based on a strong physical and engineering background, while the change-of-variable method is a pure mathematical manner. For example, without an engineering background, it is difficult to reject the transformation (14). From the pure mathematical point of view only, it is also difficult to explain why displacement mapping can only have the form of Eq. (4). Thus, deeper physical insight should be introduced to the transformation method.

## 3. Physical interpretation of the form-invariance

The objective of the transformation method is to control some physical fields by designed appropriate material, and the achieved material is usually inhomogeneous and anisotropic. Thus, if one wants to manipulate the physical fields in such media, he must in advance know the governing equations for these materials. According to the idea of the transformation method, such complex material is converted from a simple isotropic homogeneous one via mapping. In view of this, Hu et al. [18] give the physical interpretation of form-invariance: If Eq. (1) is equally used for both homogeneous and inhomogeneous materials, it is considered as form-invariant. In contrast, if Eq. (1) has a different form in inhomogeneous materials, it is considered as form-variant. For example, in classical physics, Maxwell's equations

$$\nabla \times \mathbf{E} = -\boldsymbol{\mu}\frac{\partial \mathbf{H}}{\partial t}, \quad \nabla \times \mathbf{H} = +\boldsymbol{\varepsilon}\frac{\partial \mathbf{E}}{\partial t}, \qquad (16)$$

can be equally used for both isotropic homogeneous materials and anisotropic inhomogeneous materials. Similarly, the classical acoustic wave equation

$$\nabla p = \rho \ddot{\mathbf{u}}, \quad p = \kappa \nabla \cdot \mathbf{u}, \qquad (17)$$

can also be equally used for both isotropic homogeneous materials and anisotropic inhomogeneous materials, where the mass density is assumed to have a tensor form in general. We thus say Eq. (16) and (17) are physically form-invariant.

Although some use Eq (4) for inhomogeneous materials, the inhomogeneous elastodynamics should be governed by Willis equations, which has a different form from Eq. (4), as shown in Ref. [23,24]. In addition, in the framework of the transformation method, one usually needs the transformed wave field to be smooth (especially in cloaking), while it is known for many years that for inhomogeneous materials, the wave field of Eq. (4) cannot be smooth but is like the scattered one [21]; thus, equations (4) cannot be equally used for both homogeneous and inhomogeneous materials. We therefore say that Eq. (4) is not physical form-invariant.

The essential difference between acoustics (or optics) and elastodynamics is that the former is established by first-order approximation, i.e., the variation of a quantity around a point is expressed by Taylor series expansion with higher-order terms being neglected. Then the physical model of Eq. (16) or (17) is simply based on a point, which can be used for both homogeneous and inhomogeneous materials. In contrast, elastodynamics must consider the moment effect. Thus the physical model is based on a finite area rather than a point that cannot support moment. The balance equation of angular momentum by first-order approximation is written as (without summation)

$$\sigma_{ij} + \frac{1}{2}\frac{\partial \sigma_{ij}}{\partial x_i}dx_i = \sigma_{ji} + \frac{1}{2}\frac{\partial \sigma_{ji}}{\partial x_j}dx_j, \quad i,j = 1,2,3, i \neq j, \quad (18)$$

where $dx_i$ is the size of an element. Equations (18) complete the elastodynamics equations in addition to Eq. (4). However, Eq. (18) are difficult to solve, so it can be simplified to

$$\sigma_{ij} = \sigma_{ji} \quad (19)$$

by zero-order approximations. Equation (19) is usually omitted in Navier's equation because it is included by the symmetry of classical elasticity tensor. Thus, equations (4) with symmetric elasticity tensor in fact use zero-order approximation and are proper only for slowly varying materials (or high-frequency waves). Detailed discussions on the precision of the governing equations are beyond the scope of this paper. What we want to emphasize is that in the physical interpretation, the "form-invariance" or "form-variance" of the governing equations is already known before obtaining Eq. (3).

It is helpful to recall the principle behind the transformation method. The anisotropic inhomogeneous material can be regarded as transformed from the isotropic homogeneous one. However the transformation can be arbitrary and different transformations will lead to different material, then how do we find the appropriate transformation that can be used to control the physical fields? The transformation method gives an elegant solution: it connects the material transformation to a space mapping and the local properties of the transformed material are determined by the local geometrical properties of this space mapping. This way, the material can be transformed via a certain visualized method point by point, in turn relating the physical fields to the material.

Hu et al. [25] further recognized that the space mapping is equivalent to space deformation. If Eq. (1) and Eq. (2) are established by first-order approximation, the local linear transformation must exist between them. So space deformation can also be interpreted by first-order approximation. In other words, space deformation is regarded as local affine deformation point-by-point, and the physical quantities transformations are equivalent to physical quantities deformations aroused by space deformation. Afterwards, according to continuum mechanics, the deformation gradient tensor $\mathbf{A}$ (equal to the Jacobian transformation tensor) induced by space mapping can be decomposed by a rotation (orthogonal) tensor $\mathbf{R}$ and a pure stretch tensor $\mathbf{V}$ as $\mathbf{A} = \mathbf{VR}$. Supposing at each point $\mathbf{e}'_i$ and $\lambda_i$ respectively the eigenvectors and eigenvalues of $\mathbf{V}$, there are $\mathbf{V} = \lambda_1 \mathbf{e}'_1 \mathbf{e}'_1 + \lambda_2 \mathbf{e}'_2 \mathbf{e}'_2 + \lambda_3 \mathbf{e}'_3 \mathbf{e}'_3$. The physical field $\mathbf{u}$ and material $\mathbf{C}$ in the initial space are rigidly rotated with the element to the local system $\mathbf{e}'$ of the transformed space. In the new local Cartesian system $\mathbf{e}'$ the components of the field $\mathbf{u}$ and material $\mathbf{C}$ will then be rescaled along $\mathbf{e}'_i$. Symbolically the transformations can be written as

$$\mathbf{V_q R}: \mathbf{q} \mapsto \mathbf{q}', \quad \mathbf{q} = \mathbf{u}, \mathbf{C}, \quad (20)$$

where $\mathbf{V_q}$ is the scaling tensor for the quantity $\mathbf{q}$, and has a diagonal form in the specially established frame $\mathbf{e}'$, i.e., $\mathbf{V_q} = \text{diag}[q_1, q_2, q_3]$, where $q_i$ are scaling factors to be determined. The form-invariance of Eq. (1) and (2) and differential relation between the two spaces $\partial/\partial x'_i = (1/\lambda_i)\partial/\partial x_i$ can lead to the some conditions for determining the scaling factors. In addition, one should assume that during mapping, each type of energy is conserved and there is no energy conversion. If $w(\mathbf{u}, \mathbf{C})$ and $w'(\mathbf{u}', \mathbf{C}')$ denote respectively any type of energy density in the initial and transformed spaces, the volume of an element $dv$ becomes $\lambda_1 \lambda_2 \lambda_3 dv$ during the mapping, so the energy conservation of each point leads to the following physical constraint condition

$$w(\mathbf{u}, \mathbf{C}) = w'(\mathbf{u}', \mathbf{C}')\lambda_1 \lambda_2 \lambda_3. \quad (21)$$

The energy conservation will provide other constraint conditions for the scaling factors. The constraint conditions will allow one to finally determine the transformation relations for the field and material.

This theory provides a general method to determine the transformation relations for any physical process governed by a set of PDE, if the PDEs are established by first-order approximation. Hu et al. [18] showed that the constraint conditions are not enough to determine completely the transformation relation for transformation acoustics, leaving a possibility to define them differently as found in the literature. They also show that the transformation relation is uniquely determined for transformation optics.

## 4. Transformation elastodynamics sample

The proposed theory can work well for those equations established by first-order approximation, because in these cases the equations can be used for both homogeneous and inhomogeneous materials. However, the method is not limited to those equations; the physical view has more flexibility in material design compared with usual methods such as change-of-variable method.

According to the physical interpretation of the form-invariance, the designer should in advance knows the governing equations for the inhomogeneous and anisotropic material, thus, even if the equations are not standard first-order approximations but, within some application scope, they can be used in inhomogeneous media, then form-invariance still exists. One example is elastodynamics. It is known that Eq. (4), with symmetric elasticity tensor, can not be used in inhomogeneous materials if a smooth wave field is required. However, in cases of high-frequency waves or slowly varying materials, these equations can be used approximatively (see Eq. (15)). Therefore, with this limit, one can still explore the proposed method to obtain the transformation relations of Eq. (4).

As shown in Ref. [21,26], research on high frequency (short wavelengths) has a long history in acoustic and elastodynamic wave propagation problems for various important applications, such as seismology, petroleum exploration, nondestructive ultrasound evaluation and other areas. In addition, as shown in Ref. [21], the condition of high frequency has only a relative meaning. It requires that the appropriate material parameters of the media do not vary greatly over an order of a wavelength. For example, in seismology, sometimes waves with frequency ≈ 8 Hz can be considered as high frequency. Thus, high frequency transformation elastodynamics is useful in practice.

The transformation relations are derived directly from constraint conditions and no transformation relation is pre-assumed. As the constraint equations of elastodynamics have non-unique solutions, various transformations can be obtained [19-20]. Interestingly, in the local conformal space mapping, which will lead to both isotropic transformed modulus and density, the impedance-matched condition for both S and P waves in perpendicularly incident cases exists in the proposed method [20]

$$E' = \lambda E, \quad \upsilon' = \upsilon, \quad \rho' = \rho/\lambda ,\qquad(22)$$

where $E$ and $\upsilon$ denote Young's modulus and Poisson's ratio of the media, respectively, and $\lambda$ is the local scaling factors of the conformal space mapping. Equations (22) show that for the perpendicularly incident elastic waves, the impedance-matched condition does not require the boundary between the transformed media and the background media to be fixed, in contrast to the electromagnetic waves [27,28]. Thus, we can design an approximate elastic waves cloak with isotropous homogeneous material. Let a region with a hole have a uniform magnifying

$$\mathbf{x}' = \lambda \mathbf{x} ,.\qquad(23)$$

where $\lambda > 1$ is a constant real. In this space mapping, one can have $\mathbf{B} = \mathbf{A} = $ const in Eq. (8), however condition (22) can not be derived by Eq. (8). With help of the designed cloak, the scattering aroused by the obstacle in the hole can be smaller compared to that without the cloak, because the cloak is equivalent to "compress" the obstacle size. Equations (22) guarantee that there is no scattering in the incident boundary, and the cloak is of isotropous and homogeneous. The space mapping is shown in Fig. 1. The simulation result is shown in Fig. 2, where the constant scaling factor $\lambda = 3$ and the background media is structural steel with material parameters $E = 200\text{Gpa}$, $\upsilon = 0.33$ and $\rho = 7850 \text{ kg/m}^3$. The obstacle has a material parameter $E_{obs} = 10E$ and other parameters are the same as the background. It is shown that the cloak can obviously reduce scattering. The directional elastic cloak was also discussed by Amirkhizi et al [29] with a different method.

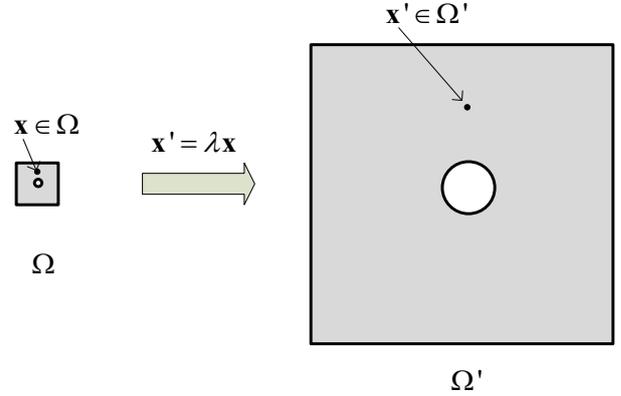

Figure 1: Space mapping of the approximate elastic waves cloak with isotropous homogeneous material.

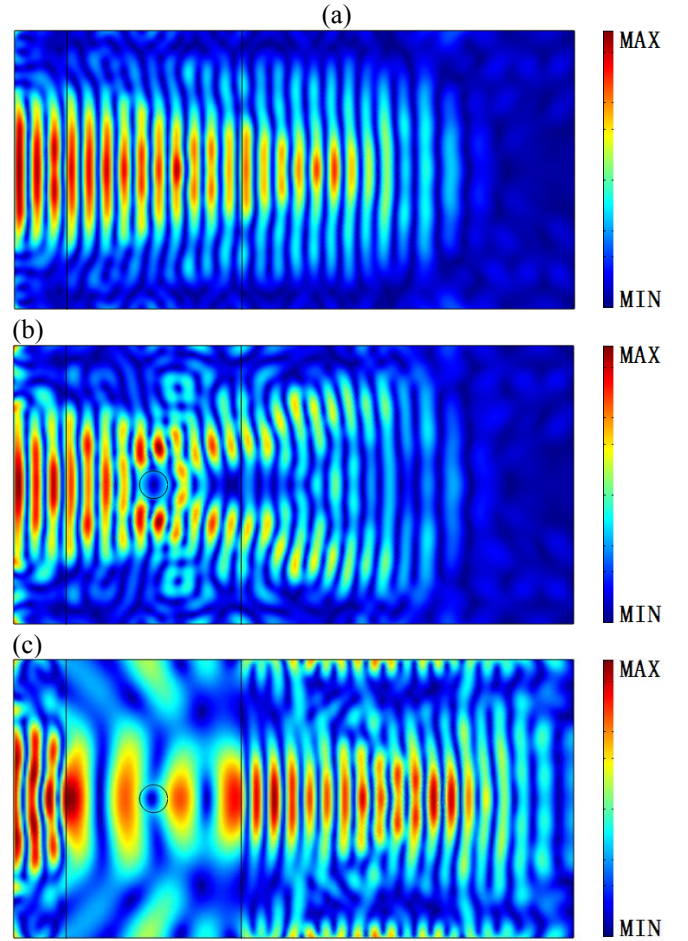

Figure 2: Simulation results (total displacement) of the designed elastic waves cloak. (a) Original waves without obstacle; (b) waves without cloak; (c) waves with the designed cloak.

## 5. Discussion and conclusions

This paper reviews some important concepts and methods contained in the transformation method. The transformation relations for field and material during a transformation are very important in this method. The usual methods to obtain Eq. (3) are based on the mathematical interpretation of the form-invariance, where the form-invariance is a pure mathematical property of an equation that one needs to verify. The change-of-variable method is a representative method of this idea. It needs to pre-assume certain transformation relations for some physical quantities, and then express the others in a general arbitrary curvilinear coordinate system. Thus the results depend on the pre-assumed transformation relations and this method cannot obtain full transformation relations. More importantly, the mathematical interpretation of form-invariance masks the physical essence implied by the transformation method, i.e., this method in fact cares about what kind of mapping the equations have, instead of whether the equations can be mapped.

A physical interpretation of form-invariance is explored. This form-invariance means that the governing equation is used for both homogeneous and inhomogeneous materials. One can interpret the general mapping by first-order approximation to guarantee the form-invariance. The transformed material property and physical field are constrained by some constraint conditions. In summary, the proposed method has these characteristics: (1) There is no need to find the transformed equation; (2) The local affine transformation is introduced to assure the same form of equations in homogeneous and inhomogeneous materials (we know they have same form from physics); (3) There are no pre-assumed transformations for the physical quantities; (4) Some physical significance of the transformation method, i.e., rigid rotation and pure stretch of the physical quantity, as well as the energy conservation, are investigated; (5) The application scope of the transformation relations are known before obtaining them, because any reasonable governing equation must have a clear answer for their application scope. The proposed theory provides a general method to determine the transformation relations for any physical process governed by a set of PDE within its application scope, and thus, we have a convenient method to explore the transformation method for a vast range of potential dynamical systems.

### Acknowledgements

This work was supported by the National Natural Science Foundation of China (11172037) and Excellent Young Scholars Research Fund of Beijing Institute of Technology.